\documentclass[aps,prd,notitlepage,superscriptaddress,showpacs,nofootinbib,10pt]{revtex4-1}
\usepackage{epsfig}
\usepackage{xcolor}
\usepackage{amsmath}
\usepackage{amssymb}
\usepackage{bm}
\usepackage{graphicx,psfrag}
\usepackage{epsfig}
\usepackage{afterpage}
\usepackage{mathrsfs}
\newcommand{\ben}{\begin{eqnarray}}
\newcommand{\een}{\end{eqnarray}}
\newcommand{\nnu}{\nonumber\\}
\newcommand{\bef}{\begin{figure}[htb]\centering}
\newcommand{\eef}{\end{figure}}
\newcommand{\sla}[1]{{#1}\!\!\!\slash}

\begin{document}
\title{Single transverse spin asymmetry of prompt photon production}

\author{Leonard Gamberg}
\email{lpg10@psu.edu}
\affiliation{Division of Science, 
                  Penn State Berks, 
                  Reading, PA 19610, USA}

\author{Zhong-Bo Kang}
\email{zkang@lanl.gov}
\affiliation{Theoretical Division, 
                   Los Alamos National Laboratory, 
                   Los Alamos, NM 87545, USA}

\begin{abstract}
We study the single transverse spin asymmetry of prompt photon production in high energy proton-proton scattering. We include the contributions from both the direct and fragmentation photons. While the
asymmetry for direct photon production receives only the Sivers type of contribution, the asymmetry for fragmentation photons receives both the Sivers and Collins types of contributions. We make a model calculation for quark-to-photon Collins function, which is then used to estimate the Collins asymmetry for fragmentation photons. We find that the Collins asymmetry for fragmentation photons is very small, thus the single transverse spin asymmetry of prompt photon production is mainly coming from the Sivers asymmetry in direct and fragmentation photons.
 We make predictions for the prompt photon spin asymmetry at RHIC energy, and emphasize the importance of such a measurement. The asymmetry of prompt photon production can  provide a good measurement for the important twist-three quark-gluon correlation function, which is urgently needed in order to resolve the ``sign mismatch'' puzzle.

\end{abstract}

\date{\today}

\maketitle

\section{Introduction}
Single spin asymmetries (SSAs) in transversely polarized proton-proton collisions have provided 
essential information on the internal partonic structure of the proton, particularly the parton's transverse motion in the transversely polarized proton \cite{Boer:2011fh}. Two different yet related QCD factorization formalisms have been proposed to describe the observed asymmetries: the transverse momentum dependent (TMD) factorization \cite{TMD-fac,Brodsky,MulTanBoe, Boer:2003cm,boermulders}  and the collinear twist-three factorization approaches \cite{Efremov,qiu,koike,Kang:2008qh,Koike:2011nx,Koike:2011mb,Kang:2010zzb}. 

For  processes such as semi-inclusive hadron production in 
lepton-proton deep inelastic scattering (SIDIS) $\ell p^\uparrow\to \ell' h X$
which are characterized by both the photon virtuality $Q^2$ and 
hadron transverse momentum  $P_{h\perp}$ such that $Q\gg P_{h\perp}\sim \Lambda_{\rm QCD}$,  one  
describes the SSAs in the TMD factorization formalism. In this approach
the transverse spin effects are associated with  naive time-reversal-odd
TMDs which represent helicity flip cut quark target scattering amplitudes
with a non-trivial color phase~\cite{Gamberg:2010uw}.   Two well-known TMDs are the quark Sivers function \cite{Siv90} and Collins function \cite{Collins93}, which describe the so-called $\sin(\phi_h -\phi_s)$ and $\sin(\phi_h +\phi_s)$ modulations in SIDIS on transversely polarized target, respectively. Because of the different angular modulations in the cross section, 
one can separate Sivers from Collins effect in SIDIS and thus extract them independently from the experimental data~\cite{:2009ti,Alekseev:2008aa,Qian:2011py}. 
On the other hand, for single inclusive hadron production in proton-proton scattering 
$p^\uparrow p\to h X$ where there is a single 
hard scale given by the hadron's transverse momentum, $P_{h\perp}\gg \Lambda_{\rm QCD}$, one can 
 describe the SSAs in the collinear twist-three factorization approach in terms of either the twist-three quark-gluon correlation functions in the transversely polarized 
proton \cite{Kouvaris:2006zy,Kanazawa:2011bg}, or the twist-three fragmentation functions in the hadronization process \cite{Kang:2010zzb}. In the twist-three  formalism 
we refer to  the former contribution as Sivers effect, and the latter one as Collins effect, since they 
represent  the collinear version of these two effects (based on the operator definitions relating
the first $k_T$-moments of the Sivers and Collins functions 
in the collinear twist-three approach~\cite{Boer:2003cm}). 
While the most abundant experimental data exist  \cite{Adams:1991cs, SSA-rhic} 
on transverse spin effects
in  SSAs of single inclusive hadron production in proton-proton collisions, 
 disentangling Sivers and Collins  contributions  presents a significant experimental challenge, 
thus the true origin (the relative contributions from these two effects) for the inclusive hadron production still remains elusive~\cite{Anselmino:2004ky}.

Theoretically it has been found that these two formalisms are closely related to each other, and it is shown that they are equivalent in the overlapping transverse momentum region where both can apply \cite{Ji:2006ub,Koike:2007dg,Bacchetta:2008xw}. However, it has been recently observed that the 
experimental proton-proton data on the SSAs 
of the inclusive hadron production appears incompatible with the Sivers 
data from SIDIS process \cite{Kang:2011hk,Kang:2012xf,Gamberg:2010tj}, if one assumes that 
the SSAs of the inclusive hadron production come entirely from the Sivers contribution. This is  known as the ``sign mismatch''. Whether this finding reflects the inconsistency of our theoretical formalism  is a very important question and needs to be further explored both theoretically and experimentally. Since the inclusive hadron production  has the complication from the Collins contribution, the measurement for the SSAs of single inclusive jet and direct photon production in proton-proton collisions \cite{Kang:2011hk, D'Alesio:2010am} could be very helpful in studying the sign mismatch, as they are free of complication from the fragmentation process (or the Collins effect).

Even though direct photon production is ideal in the theoretical sense for further exploring this  ``sign mismatch'', there are no true direct photons. Direct photons and fragmentation photons are two indistinguishable contributions in the usual collinear factorization formalism \cite{Gordon:1993qc}, which are designated  ``prompt'' photons. In experiments one might apply the photon isolation cut to reduce the fragmentation contribution, however the asymmetry measurement might suffer from the low photon event rates after such a cut. In any case, it is important to assess how the fragmentation contribution might affect the asymmetry of the prompt photons. This is the main purpose of our letter. While the direct photons receive only the Sivers type of contribution for the asymmetry,  the fragmentation photons could receive both the Sivers and Collins contributions. We perform a model calculation for the quark-to-photon Collins function, which is then used to estimate the Collins asymmetry for fragmentation photons. 

The rest of our letter is organized as follows. In Sec. II, we give the overview on the various sources for the SSAs of prompt photon production. In Sec. III, we present our detailed model calculation for the quark-to-photon unpolarized fragmentation function and Collins function, and estimate their relative size. In Sec. IV, we make phenomenological study for the SSAs of prompt photon production by including all the sources studied in our letter. We conclude our paper in Sec. V.

\section{Single transverse spin asymmetry of prompt photon production}
\subsection{Unpolarized prompt photon production}
We consider the prompt photon production in hadronic collisions, $A(P_A, s_\perp)+B(P_B)\to \gamma(P_\gamma)+X$. Here $A$ is a transversely polarized proton with spin vector $s_\perp$, and  $B$ is an unpolarized proton. The spin-averaged differential cross section of prompt photon production contains both direct and fragmentation contributions, 
\ben
E_\gamma\frac{d\sigma}{d^3P_\gamma}=E_\gamma\frac{d\sigma^{\rm dir}}{d^3P_\gamma}+E_\gamma\frac{d\sigma^{\rm frag}}{d^3P_\gamma}.
\label{avg}
\een
At leading order, the direct contribution is given by
\ben
E_\gamma\frac{d\sigma^{\rm dir}}{d^3P_\gamma}=
\frac{\alpha_{\rm em} \alpha_s}{s}\sum_{a,b} 
\int \frac{dx'}{x'}f_{b/B}(x') \int \frac{dx}{x} f_{a/A}(x)
H^U_{ab\to \gamma}(\hat s,\hat t,\hat u)\delta\left(\hat s+\hat t+\hat u\right),
\een
where $s=(P_A+P_B)^2$, $f_{a/A}(x)$ and $f_{b/B}(x')$ are the spin-averaged parton distribution functions, $\hat s$, $\hat t$, and $\hat u$ are the usual Mandelstam variables at the parton level. 
$H^U_{ab\to \gamma}$ are the well-known partonic hard-scattering functions for direct photon production \cite{Owens:1986mp, Kang:2011rt}. At the leading order, they are calculated from the partonic channels $qg\to \gamma q$ and $q\bar{q}\to \gamma g$, and the typical Feynman diagrams are shown in Fig.~\ref{direct}.
\bef
\psfig{file=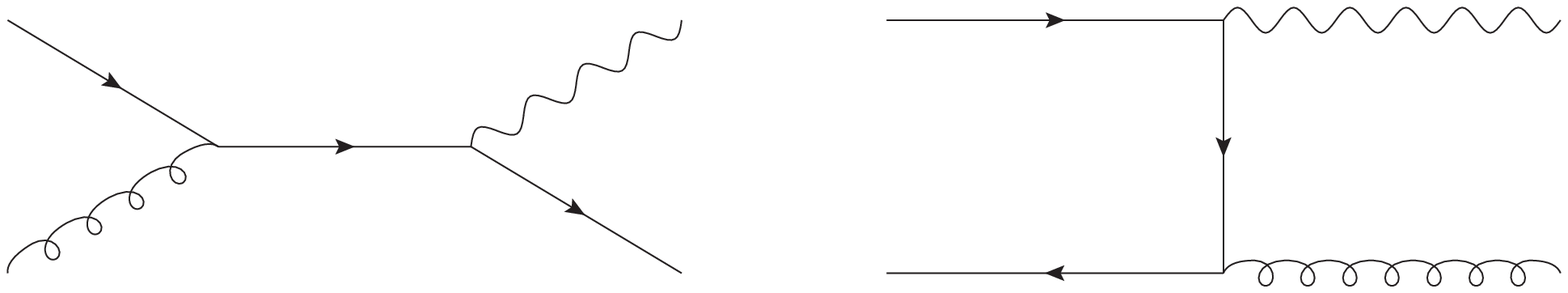,width=3.2in}
\caption{Typical Feynman diagrams for direct photon production at leading order: left for $qg\to \gamma q$ and right for $q\bar{q}\to \gamma g$.}
\label{direct}
\eef

For fragmentation photons, in the usual collinear factorization formalism at leading order, we have $2\to 2$ scattering process to produce a parton which then fragments into a photon, with the typical Feynman diagrams shown in Fig.~\ref{frag}. The differential cross section is given by
\ben
E_\gamma\frac{d\sigma^{\rm frag}}{d^3P_\gamma}=
\frac{\alpha_s^2}{s}\sum_{a,b,c} \int \frac{dz}{z^2} D_{c\to \gamma}(z)
\int \frac{dx'}{x'}f_{b/B}(x')\int \frac{dx}{x} f_{a/A}(x)
H^U_{ab\to c}(\hat s,\hat t,\hat u)
\delta\left(\hat s+\hat t+\hat u\right),
\een
where $D_{c\to \gamma}(z)$ is the quark-to-photon fragmentation function, and $H^U_{ab\to c}$ are the 
well-known partonic cross section to produce a parton \cite{Owens:1986mp, Kang:2011bp}.
\bef
\psfig{file=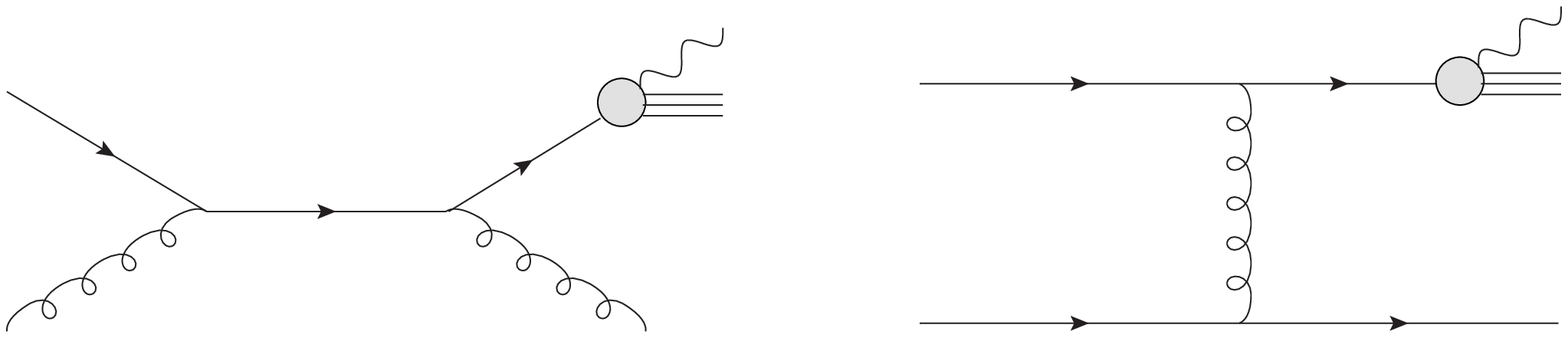,width=3.2in}
\caption{Typical Feynman diagrams for fragmentation photon production at leading order.}
\label{frag}
\eef

To see the relative contributions of direct and fragmentation photons, we define the following direct ratio
\ben
R=\frac{E_\gamma\frac{d\sigma^{\rm dir}}{d^3P_\gamma}}{E_\gamma\frac{d\sigma^{\rm dir}}{d^3P_\gamma}+E_\gamma\frac{d\sigma^{\rm frag}}{d^3P_\gamma}}.
\label{ratio}
\een
In Fig.~\ref{figratio}, we plot the direct ratio $R$ as a function of Feynman $x_F$ at forward rapidity $y=3.5$ at RHIC energy $\sqrt{s}=200$ GeV. We give the result for both the leading order and next-to-leading order calculations \cite{Gordon:1993qc}. We find that the fragmentation photons actually contributes to around $50\%$ to the total prompt photon production. Thus it is important to assess the effect of fragmentation photons on the asymmetry of the prompt photon production. 
\bef
\psfig{file=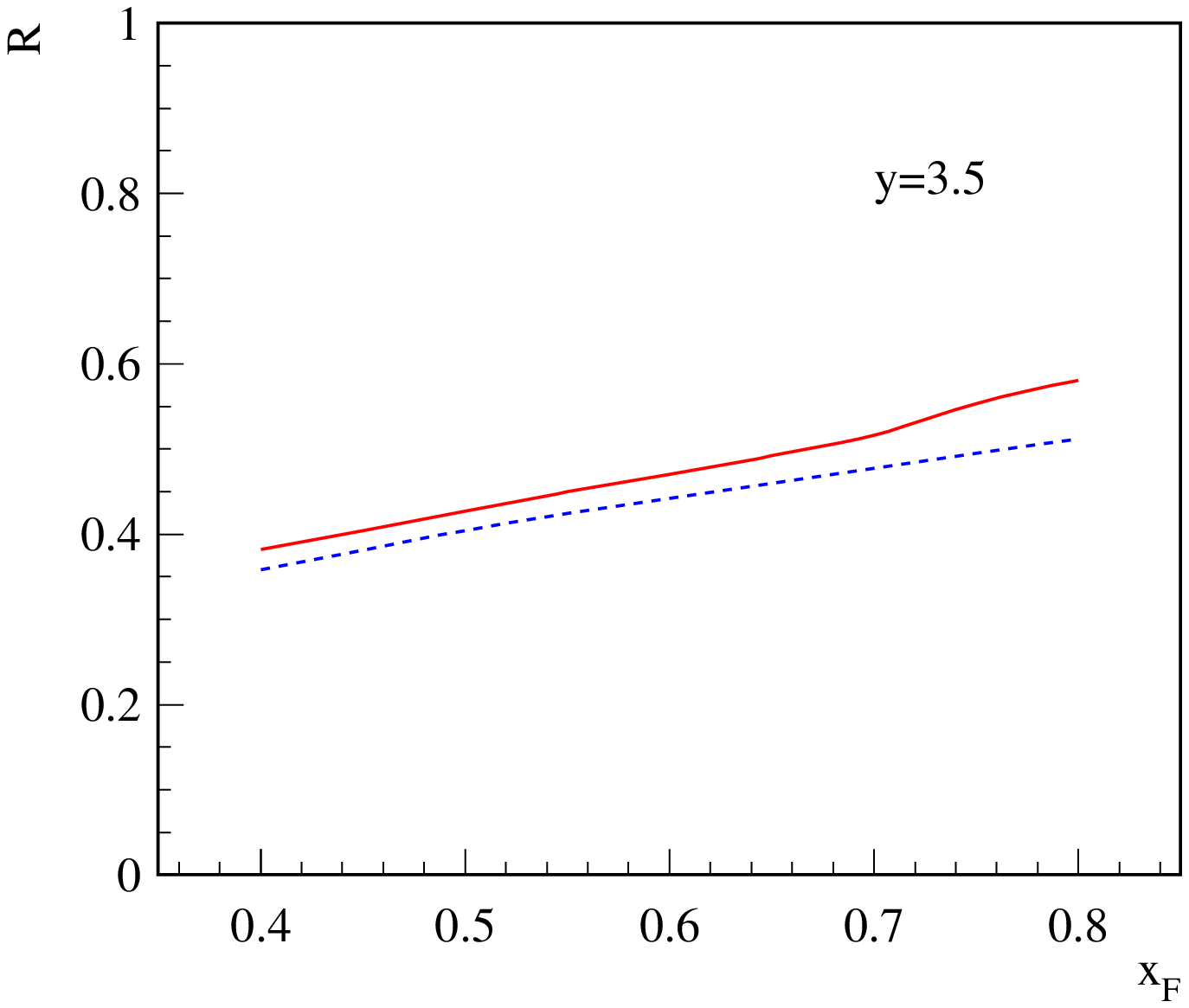, width=2.8in}
\caption{The direct ratio defined in Eq.~(\ref{ratio}) is plotted as a function of Feynman $x_F$ at $y=3.5$ and $\sqrt{s}=200$ GeV. The solid line is for leading order calculation, while the dashed line is for next-to-leading order calculation.}
\label{figratio}
\eef

\subsection{Spin-dependent cross section for prompt photon production}
In order to compute the asymmetry of prompt photon production, we need the spin-dependent cross section $\Delta\sigma(s_\perp) = \left[\sigma(s_\perp) - \sigma(-s_\perp)\right]/2$, which will also contain both direct and fragmentation contributions,
\ben
E_\gamma\frac{d\Delta\sigma}{d^3P_\gamma}=E_\gamma\frac{d\Delta\sigma^{\rm dir}}{d^3P_\gamma}+E_\gamma\frac{d\Delta\sigma^{\rm frag}}{d^3P_\gamma}.
\label{davg}
\een
The direct contribution contains only the Sivers type of effect, as given by \cite{Kouvaris:2006zy,Kang:2011hk}
\ben
E_\gamma\frac{d\Delta\sigma^{\rm dir}}{d^3P_\gamma} &=& 
\epsilon_{\alpha \beta} s_\perp^\alpha P_{\gamma\perp}^\beta
\frac{\alpha_{\rm em}\alpha_s}{s}\sum_{a,b}
\int \frac{dx'}{x'}f_{b/B}(x')\int \frac{dx}{x}
\left[T_{a,F}(x, x) - x\frac{d}{dx}T_{a,F}(x, x)\right]
\nnu
&&
\times
\frac{1}{\hat{u}} H^{\rm dir}_{ab\to \gamma}(\hat s,\hat t,\hat u)\delta\left(\hat s+\hat t+\hat u\right),
\een
where the hard-part functions $H^{\rm dir}_{ab\to \gamma}$
contain the relevant initial-state interactions between the active parton and the remnant of the proton and have the expressions given in \cite{Gamberg:2010tj,Kouvaris:2006zy}. $T_{q,F}(x, x)$ is the twist-three quark-gluon correlation function, and it is related to the quark Sivers function $f_{1T}^{\perp q}(x, k_\perp^2)$ as follows~\cite{Boer:2003cm}
\ben
T_{q, F}(x, x)=-\int d^2k_\perp \frac{|k_\perp|^2}{M}f_{1T}^{\perp q}(x, k_\perp^2)|_{\rm SIDIS},
\label{TF}
\een
where the subscript ``SIDIS'' here is to emphasize the Sivers function probed in SIDIS process.  On the other hand, the spin asymmetry of fragmentation photons can receive both Sivers and Collins contributions,
\ben
E_\gamma\frac{d\Delta\sigma^{\rm frag}}{d^3P_\gamma} = E_\gamma\frac{d\Delta\sigma^{\rm frag}_{\rm Sivers}}{d^3P_\gamma}+ E_\gamma\frac{d\Delta\sigma^{\rm frag}_{\rm Collins}}{d^3P_\gamma}.
\een
The Sivers contribution can be written as
\ben
E_\gamma\frac{d\Delta\sigma^{\rm frag}_{\rm Sivers}}{d^3P_\gamma} &=& 
\epsilon_{\alpha \beta} s_\perp^\alpha P_{\gamma\perp}^\beta
\frac{\alpha_s^2}{s}
\sum_{a,b,c} \int \frac{dz}{z^2} D_{c\to \gamma}(z)
\int \frac{dx'}{x'}f_{b/B}(x')\int \frac{dx}{x}
\left[T_{a,F}(x, x) - x\frac{d}{dx}T_{a,F}(x, x)\right]
\nnu
&&
\times
\frac{1}{z \hat{u}}
H^{\rm Sivers}_{ab\to c}(\hat s,\hat t,\hat u)\delta\left(\hat s+\hat t+\hat u\right),
\een
where $H^{\rm Sivers}_{ab\to c}$ represents a hard-part functions for the partonic process $a b\to c d$, and it incorporates both the initial and final state interactions and has the expressions given in \cite{Kouvaris:2006zy,Kang:2011hk}. The Collins contribution for an inclusive hadron production has been calculated in \cite{Kang:2010zzb}, which is related to a convolution of quark transversity and quark-to-hadron twist-three fragmentation function. The only difference for fragmentation photons lies in the quark-to-photon twist-three fragmentation function, and the differential cross section is given by
\ben
E_\gamma\frac{d\Delta\sigma^{\rm frag}_{\rm Collins}}{d^3P_\gamma}&=&
\epsilon_{\alpha \beta} s_\perp^\alpha P_{\gamma\perp}^\beta
\frac{ \alpha_s^2}{s}
\sum_{a,b,c}
\int \frac{dx}{x} h_a(x) \int \frac{dx'}{x'}f_b(x') \int \frac{dz}{z}
\left [ -z \frac{\partial}{\partial z}\left(\frac{\hat{H}_c(z)}{z^2}\right) \right ]\nonumber \\
&&\times
\left[\frac{1}{z}\frac{x-x'}{x(-\hat u)+x' (-\hat t)}\right]
H^{\rm Collins}_{ab \rightarrow c}( \hat{s}, \hat{t}, \hat{u} ) 
\delta\left(\hat s+\hat t+\hat u\right)\, ,
\een
where $h_a(x)$ is the quark transversity, and $\hat H_c(z)$ is the twist-three quark-to-photon fragmentation function and is related to the first $p_T$-moment of the Collins function $H_1^{\perp q}(z, p_T^2)$
\ben
\hat{H}_q(z)=-\frac{1}{z}\int d^2p_T\, p_T^2\, H_1^{\perp q}(z, p_T^2),
\label{hqz}
\een
with $H_1^{\perp q}(z, p_T^2)$ defined in the next section. The relevant hard-part function $H^{\rm Collins}_{ab \rightarrow c}$ has been computed in \cite{Kang:2010zzb}.

Eventually the single transverse spin asymmetry $A_N$ is computed from the following definition
\ben
A_N = \frac{E_\gamma\frac{d\Delta\sigma}{d^3P_\gamma}}{E_\gamma\frac{d\sigma}{d^3P_\gamma}},
\label{ssa}
\een
where the spin-dependent and spin-averaged cross sections are given in Eqs.~(\ref{avg}) and (\ref{davg}), respectively. To calculate $A_N$ numerically, we need the information for the twist-three quark-gluon correlation function $T_{q, F}(x, x)$ and twist-three quark-to-photon fragmentation function $\hat H_q(z)$. The information of $T_{q, F}(x, x)$ has been directly extracted from the proton-proton data \cite{Kouvaris:2006zy, Kanazawa:2011bg}, or indirectly from the SIDIS data by using Eq.~(\ref{TF}) \cite{Anselmino:2005ea,Anselmino:2008sga}. However, the information of $\hat H_q(z)$ is completely unknown. To estimate the size of $\hat H_q(z)$ will be the main focus of the next section.

\section{quark to photon Collins function}\label{sec3}
In this section, we perform model calculations for photon fragmentation functions, including both the unpolarized fragmentation function and the Collins function. We first study the transverse momentum dependent quark-to-photon fragmentation functions, and then integrate over the transverse momentum to obtain the relevant unpolarized collinear fragmentation function $D_{q\to \gamma}(z)$ and the collinear twist-three fragmentation function $\hat{H}_q(z)$.

\subsection{Transverse momentum dependent fragmentation functions}\label{sec3a}
Photon fragmentation function can be calculated from the correlation function $\Delta(z, k_T)$ 
\cite{MulTanBoe,Boer:2003cm,Bacchetta:2000jk},
\ben
\Delta(z, k_T) = \frac{1}{2z} \sum_X \int \frac{d\xi^+ d^2\xi_T}{(2\pi)^3} e^{ik\cdot \xi} 
\langle 0 | \psi_q(\xi)|\gamma X\rangle \langle \gamma X| \bar\psi_q(0)|0\rangle|_{\xi^-=0},
\een
where the usual gauge link is suppressed, and we have assumed that the photon is moving in $-z$ direction with momentum $p^\mu = p^- n^\mu$ and light-cone vector $n^\mu= [0^+, 1^-, 0_\perp]$. The fragmenting quark has momentum $k$, with $k^- = p^- /z$ and $k_T$ the transverse component with respect to the photon momentum $p$. We define $p_T$ as the photon transverse momentum with respect to the quark, which is related to $k_T$ as: $\vec{p}_T = -z \vec{k}_T$.  
Here for our purpose we only keep the terms relevant to the 
quark to {\it unpolarized} photon fragmentation. Then the correlation function $\Delta(z, k_T)$ is given by~\cite{MulTanBoe, Boer:2003cm,Bacchetta:2004jz},
\ben
\Delta(z, k_T)= \frac{1}{2}\left[D_{q\to \gamma}(z, p_T^2) \sla{n} + H_1^{\perp q}(z, p_T^2) \sigma^{\mu\nu}k_{T\mu}n_{\nu}\right].
\een
$D_{q\to \gamma}(z, p_T^2)$ is 
the usual unpolarized quark-to-photon fragmentation function, 
and $H_1^{\perp q}(z, p_T^2)$ is the quark-to-photon Collins function 
in agreement with the ``Trento conventions'' \cite{Bacchetta:2004jz}.
For the most general 
case where photon's polarization is also specified, there are more terms 
in the expansion \cite{Bacchetta:2000jk}.  
 We can easily project out these functions
\ben
D_{q\to \gamma}(z, p_T^2) &=& \frac{1}{2} {\rm Tr}\left[\Delta(z, k_T) \sla{\bar{n}}\right],
\\
\epsilon_T^{\mu\nu} k_{T\nu} H_1^{\perp q}(z, p_T^2) &=&  \frac{1}{2} {\rm Tr} \left[\Delta(z, k_T)  i \sigma^{\mu\nu}\bar{n}_\nu\gamma^5\right],
\een
where $\bar{n}^\mu= [1^+, 0^-, 0_\perp]$ is a light-cone vector conjugate to $n^\mu$. 

\bef
\psfig{file=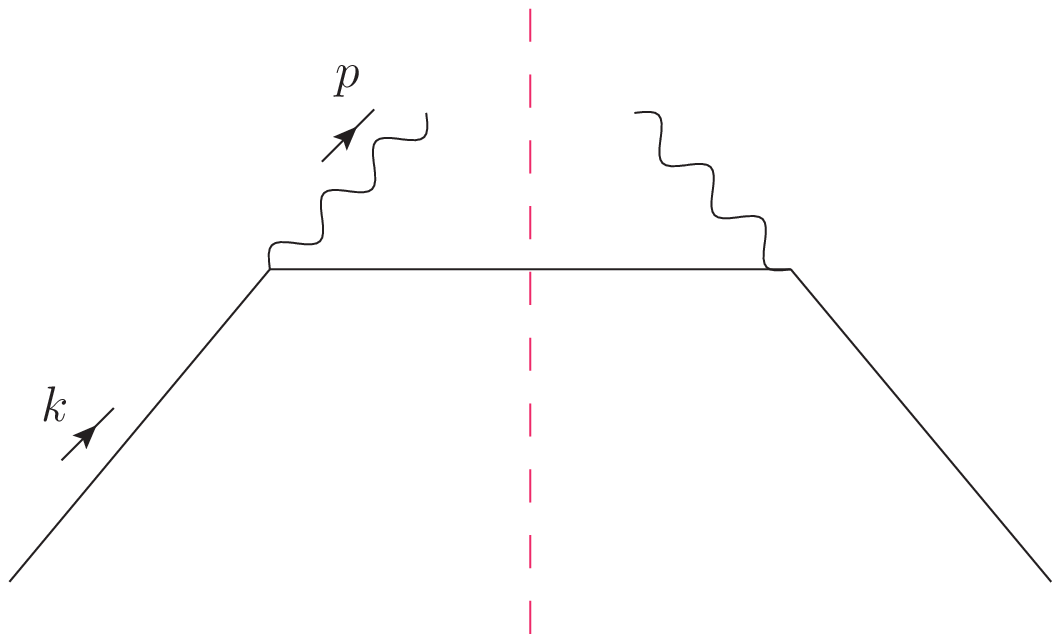, width=2in}
\caption{The Feynman diagram which contribute to the unpolarized quark-to-photon fragmentation function $D_{q\to \gamma}(z, p_T^2)$.}
\label{unp}
\eef
In our  model, the tree-level diagram describing the fragmentation of a quark into a real photon is depicted in Fig.~\ref{unp}.  By contrast with the pion fragmentation calculations~\cite{Gamberg:2003eg,Amrath:2005gv,Bacchetta:2007wc}, the interaction between quark and the photon is described by a simple point interaction with coupling $ie_q e \gamma^\mu$ and $e_q$ the quark fractional charge. In the actual calculations, we will choose light-cone gauge $\bar{n}\cdot A_{\rm em}=0$ for the photon field \cite{Qiu:2001nr} to avoid photon eikonal phase \cite{Braaten:2001sz}. On the other hand, we still use the covariant gauge for the gluon field, thus eikonal phase for gluon field still exists in our calculations. 
In such a set-up, we have only one Feynman diagram (at leading order) for unpolarized quark-to-photon fragmentation function $D_{q\to \gamma}(z, p_T^2)$, as shown in Fig.~\ref{unp}. Thus, the photon polarization sum is given by
\ben
\sum_\lambda \epsilon^\mu(p, \lambda)  \epsilon^{*\nu}(p, \lambda) = -g^{\mu\nu} + \frac{p^\mu \bar{n}^\nu + p^\nu \bar{n}^\mu}{\bar{n}\cdot p}.
\een
The calculation is straightforward, and we obtain
\ben
D_{q\to \gamma}(z, p_T^2) = e_q^2 \frac{\alpha_{\rm em}}{2\pi^2} \frac{1}{z^2(1-z)} \left[\frac{1+(1-z)^2}{k^2-m_q^2}-\frac{2 z m_q^2}{(k^2 - m_q^2)^2}\right],
\een
where $\alpha_{\rm em}$ is the electro-magnetic coupling constant, $m_q$ is the quark mass. $k^2$ is the virtuality of the fragmenting quark, and it is related to photon transverse momentum $p_T$ as follows:
\ben
k^2 = \frac{p_T^2}{z(1-z)} + \frac{m_q^2}{1-z}.
\label{k2}
\een

\bef
\psfig{file=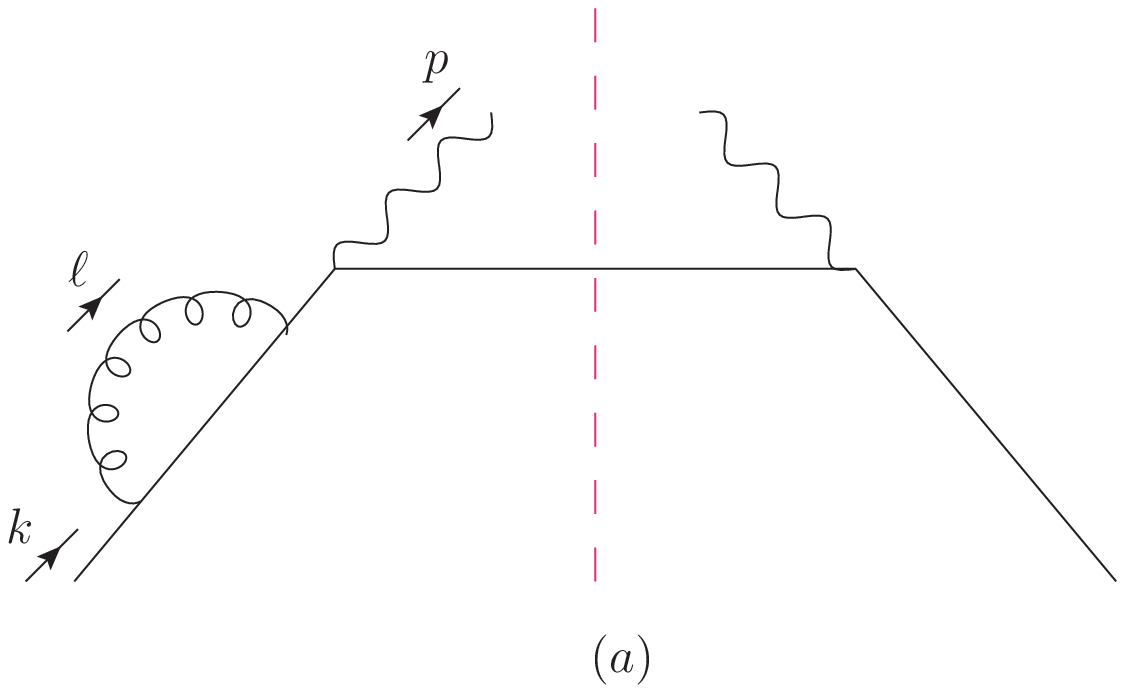,width=2in}
\hskip 0.3in
\psfig{file=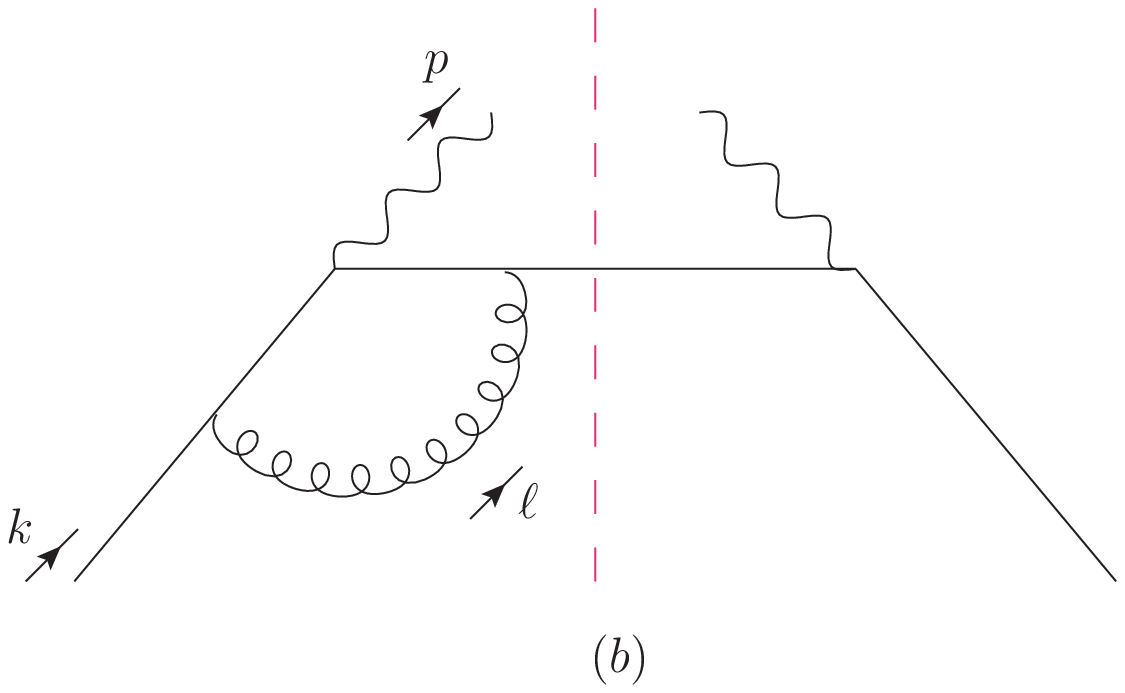,width=2in}
\\
\vskip 0.2in
\psfig{file=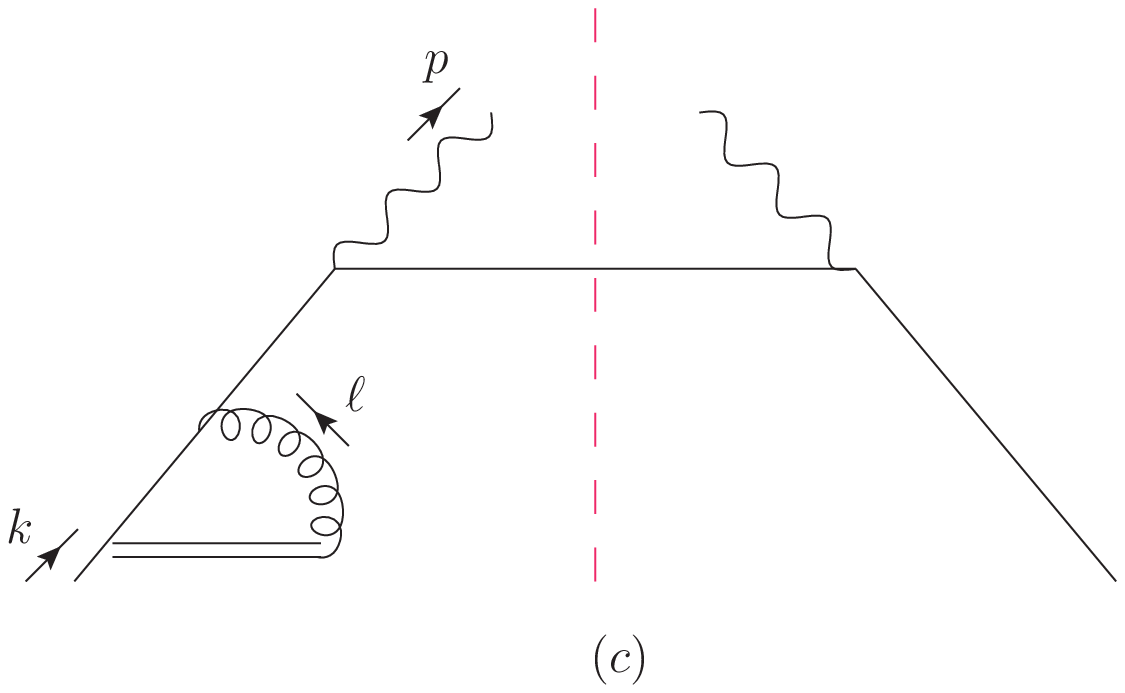,width=2in}
\hskip 0.3in
\psfig{file=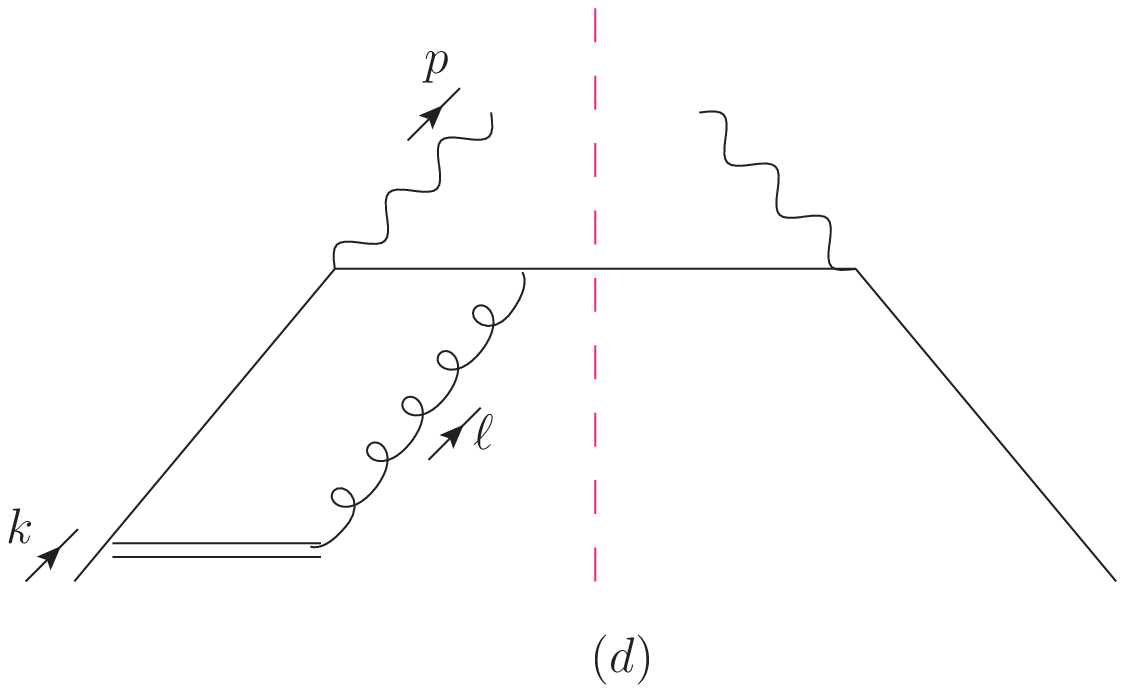,width=2in}
\caption{The Feynman diagrams which contribute to the quark-to-photon Collins fragmentation function  $H_1^{\perp q}(z, p_T^2)$. The mirror diagrams with the gluon in the right-hand side of the cut are not shown here, but are included in the calculations.}
\label{collins}
\eef

The Collins function receives contributions only from the interference between two amplitudes with different imaginary parts. Since the tree-level amplitude is real, the necessary imaginary parts will be generated by the inclusion of one-loop corrections. Here we study the case of gluon loops~\cite{Metz:2002iz,Gamberg:2003eg,Amrath:2005gv,Bacchetta:2007wc}.  The relevant Feynman diagrams are given by Fig.~\ref{collins}. The double line in Fig.~\ref{collins}(c) and (d) represents the eikonalized propagator, which give rise to the factor $1/(-\bar{n}\cdot \ell \pm i\epsilon)$ \cite{Amrath:2005gv, Bacchetta:2007wc}. The calculations are  much more involved than the unpolarized fragmentation function, but nevertheless similar to those for the quark-to-pion Collins functions calculated in \cite{Amrath:2005gv,Bacchetta:2007wc}. In particular we note that the contribution from Fig.~\ref{collins}(d) are due to poles on the gluon and  
incoming quark~\cite{Metz:2002iz,Amrath:2005gv,Bacchetta:2007wc}  signaling that the photon 
Collins function is universal~\cite{Metz:2002iz,Collins:2004nx}. Here we give only the final results,
\ben
H_1^{\perp q}(z, p_T^2) = e_q^2 \frac{\alpha_{\rm em}}{2\pi^2} \frac{m_q}{k^2-m_q^2} \alpha_s  C_F \left[
H_1^{\perp({\rm fig.a})}+H_1^{\perp({\rm fig.b})}+H_1^{\perp({\rm fig.c})}+H_1^{\perp({\rm fig.d})}
\right],
\label{gamma_Coll}
\een
where the four terms in the bracket correspond to the four diagrams in Fig.~\ref{collins} and they are given by
\ben
H_1^{\perp({\rm fig.a})}  &=& \frac{1}{2z k^2}\left(3-\frac{m_q^2}{k^2}\right),
\\
H_1^{\perp({\rm fig.b})} &=& -\frac{1}{(1-z)(k^2-m_q^2)} \left[ \frac{m_q^2}{k^2-m_q^2}\ln\left(\frac{k^2}{m_q^2}\right)
+\frac{1}{2z} \left(4-5z+3(z-2)\frac{m_q^2}{k^2}+2\frac{m_q^4}{(k^2)^2}\right)
\right],
\\
H_1^{\perp({\rm fig.c})} &=& 0,
\\
H_1^{\perp({\rm fig.d})} &=& -\frac{1}{(1-z)k^2}
\left[1+\frac{(1-z)k^2}{(1-z)k^2 - m_q^2} \ln\left(\frac{(1-z)k^2}{m_q^2}\right)\right].
\een
We note, due to the fundamental quark-photon and quark-gluon interactions
that describe the photon Collins function,
we find that the overall strength of the  various contributions in Eq.~(\ref{gamma_Coll}) are set by both
the  electro-magnetic and strong coupling. 
Moreover, we also find 
that the  function  vanishes if the quark mass is 
zero; this is consistent with the chiral-odd property 
of the Collins function~\cite{Collins93}.\footnote{Similar  quark mass 
dependence was observed for pion fragmentation in both partonic and  
effective quark-hadron model calculation of the Collins 
effect~\cite{Amrath:2005gv,Bacchetta:2007wc,Bacchetta:2002tk}.} One 
might expect such behavior in any 
partonic  model description of the photon Collins function.

\subsection{Collinear fragmentation functions}
The collinear (integrated) unpolarized fragmentation function $D_{q\to \gamma}(z)$ is defined as
\ben
D_{q\to \gamma}(z) = \pi \int_0^{p_{T\, \rm max}^2} dp_T^2\, D_{q\to \gamma}(z, p_T^2).
\een
Following \cite{Qiu:2001nr, Amrath:2005gv, Kang:2008wv}, we take the upper limit $p_{T\, \rm max}^2$ to be set by a cut-off on the fragmenting quark virtuality $\mu^2$, where  $k^2<\mu^2$. From Eq.~(\ref{k2}), this corresponds to 
\ben
p_{T\, \rm max}^2 = z(1-z) \mu^2 - z m_q^2.
\een
Then the  analytic result for $D_{q\to \gamma}(z, \mu^2)$ is
\ben
D_{q\to \gamma}(z, \mu^2) = e_q^2 \frac{\alpha_{\rm em}}{2\pi} 
\left[\frac{1+(1-z)^2}{z} \ln\frac{(1-z)(\mu^2-m^2)}{zm^2}+2\left(\frac{m^2}{\mu^2-m^2} - \frac{1-z}{z}\right)
\right].
\label{Dq}
\een

We choose a quark mass of $m_q=300$ MeV, for $D_{q\to \gamma}$  which gives 
a reasonable  estimate for quark-to-photon fragmentation function extracted from phenomenology~\cite{Gluck:1992zx} as indicated in the left panel of Fig.~\ref{hqfig}. Choosing such a mass value enables us to 
estimate the possible size  of the Collins effect to prompt photon production.
We comment more on this in  Section~\ref{pheno}.

\bef
\psfig{file=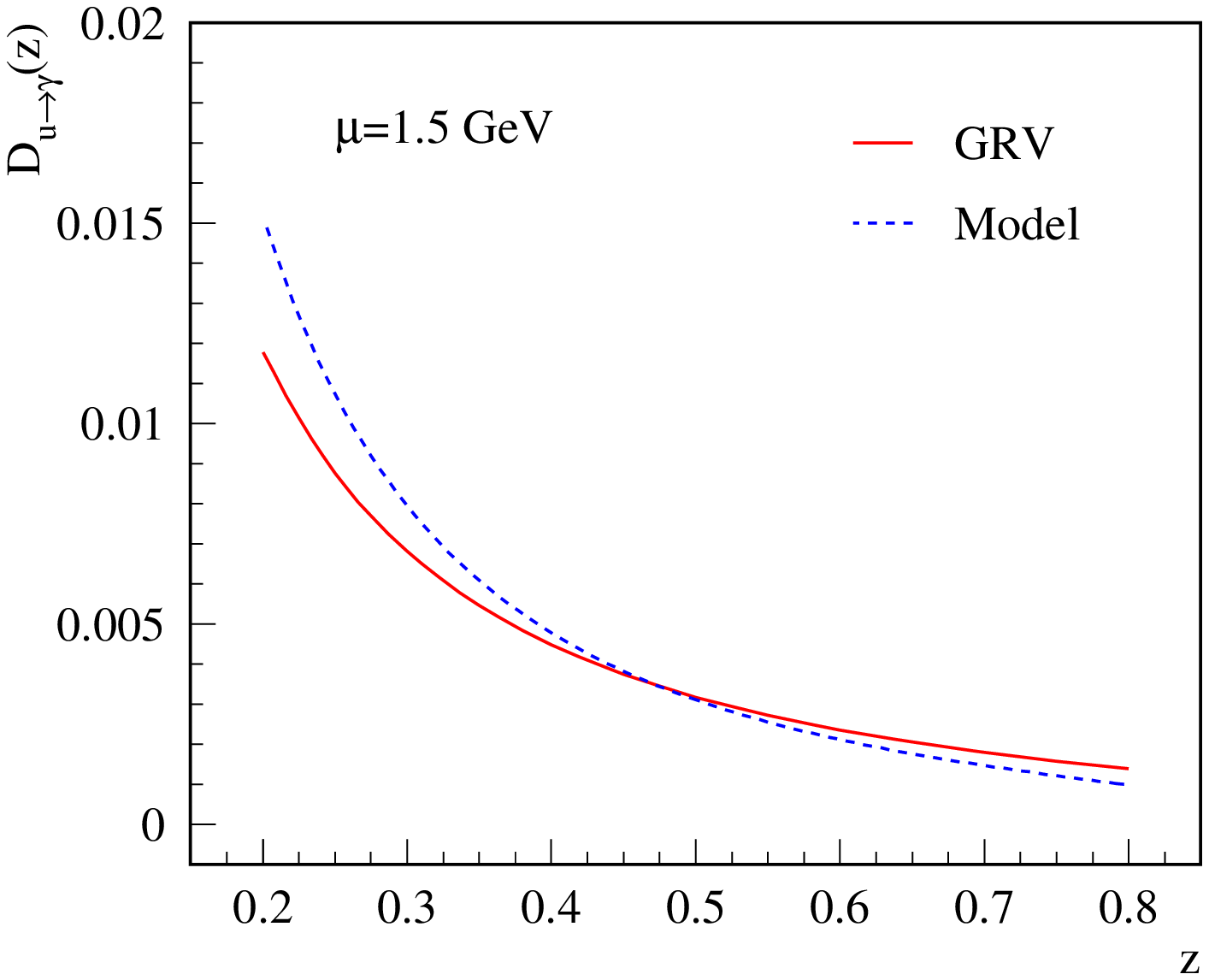, width=3.2in}
\hskip 0.3in
\psfig{file=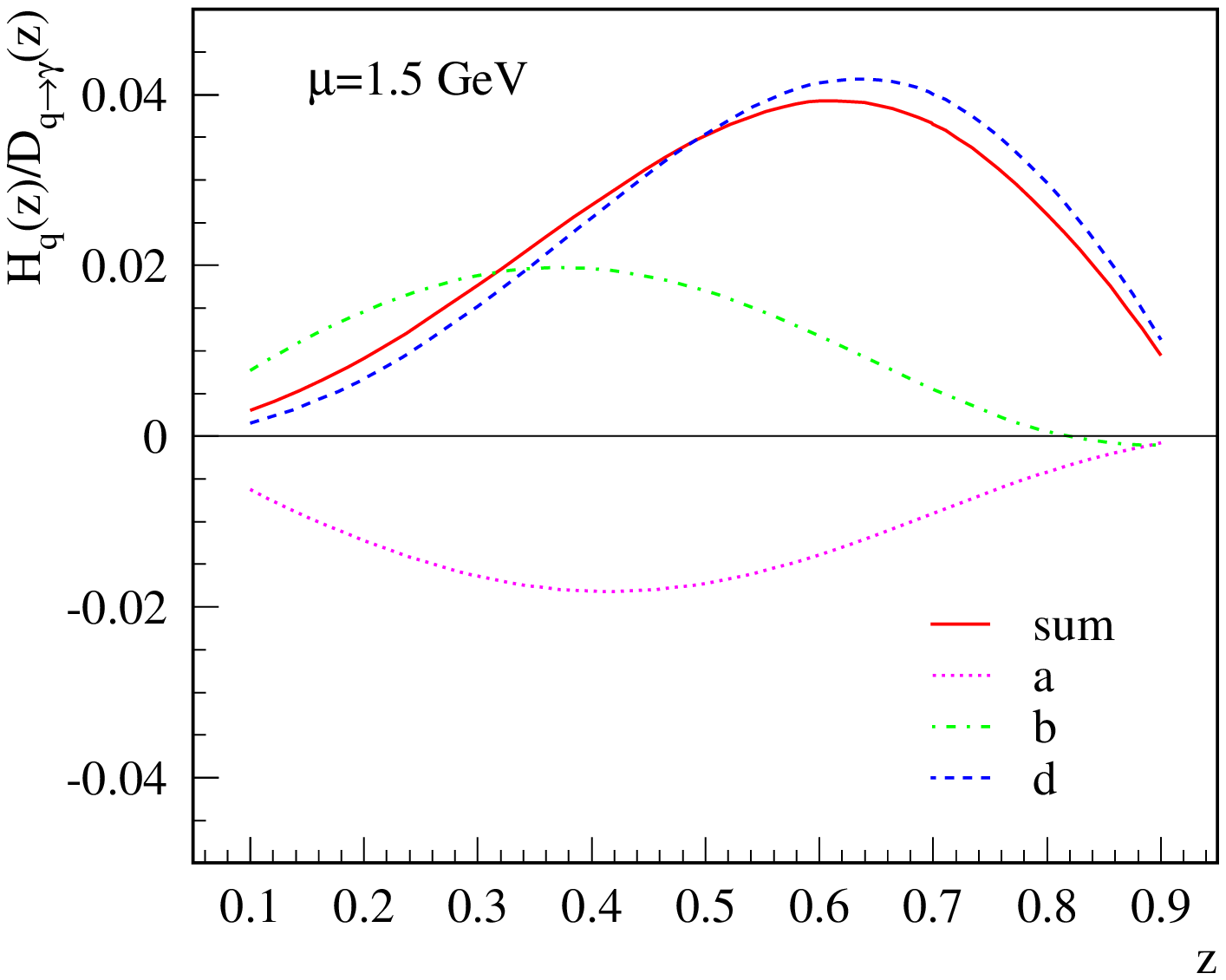,width=3.2in}
\caption{Left panel: $u$-quark to photon fragmentation function calculated from our model (blue dashed curve) in Eq.~(\ref{Dq}) with $m_q=300$ MeV compared with that extracted from phenomenology in \cite{Gluck:1992zx} (red solid curve) at $\mu=1.5$ GeV.
Right panel: The ratio of $\hat{H}_q(z, \mu^2)/D_{q\to \gamma}(z, \mu^2)$ at scale $\mu=1.5$ GeV is plotted as a function of $z$. The magenta dotted curve is the contribution from Fig.~\ref{collins}(a), the green dot-dashed for Fig.~\ref{collins}(b), the blue dashed for Fig.~\ref{collins}(d), and the red solid curve is the sum.}
\label{hqfig}
\eef
\noindent
Similarly from Eq.~(\ref{hqz}), we define the twist-three fragmentation function $\hat{H}_q(z, \mu^2)$ as 
\ben
\hat{H}_q(z, \mu^2)=-\frac{\pi}{z}\int_0^{p_{T\, \rm max}^2} dp_T^2\, p_T^2\, H_1^{\perp q}(z, p_T^2).
\label{hqcoll}
\een

Now let us estimate the relative size of twist-three fragmentation function $\hat{H}_q(z, \mu^2)$ compared to the unpolarized fragmentation function $D_{q\to \gamma}(z, \mu^2)$. 

In Fig.~\ref{hqfig} (right panel), we present numerical estimates for the analyzing power $\hat{H}_q(z, \mu^2)/D_{q\to \gamma}(z, \mu^2)$, separately for each of the diagrams of Fig.~\ref{collins} at scale $\mu=1.5$ GeV as a function of $z$.  
 The magenta dotted curve is the contribution from Fig.~\ref{collins}(a), the green dot-dashed for Fig.~\ref{collins}(b), the blue dashed for Fig.~\ref{collins}(d), and the red solid curve is the sum. We find that there is a strong cancellation between the contribution of diagrams (a) and (b), similar to the quark-to-pion Collins function \cite{Amrath:2005gv}. Thus the sum is dominantly given by the contribution from diagram (d), the gauge box diagram. We also notice that the quantity $\hat{H}_q(z, \mu^2)/D_{q\to \gamma}(z, \mu^2)$ for photon case is much smaller than the same quantity for pion case as estimated in Ref.~\cite{Amrath:2005gv}. This leads to a much smaller Collins asymmetry for fragmentation photon production, as shown in the next section. 

\section{Phenomenology}\label{pheno}
In this section, we will estimate the SSAs of the prompt photon production in the forward rapidity region at RHIC energy. In order to assess the contributions from the fragmentation photons, beside the overall spin asymmetry $A_N$ defined in Eq.~(\ref{ssa}), we will define the following additional asymmetries: the spin asymmetry for direct photon production $A_N^{\rm dir}$, the spin asymmetry for fragmentation photons
$A_N^{\rm frag}$. That is, 
\ben
A_N^{\rm dir} = \frac{E_\gamma\frac{d\Delta\sigma^{\rm dir}}{d^3P_\gamma}}{E_\gamma\frac{d\sigma^{\rm dir}}{d^3P_\gamma}},
\qquad
A_{N}^{\rm frag} = \frac{E_\gamma\frac{d\Delta\sigma^{\rm frag}_{\rm Sivers}}{d^3P_\gamma}}{E_\gamma\frac{d\sigma^{\rm frag}}{d^3P_\gamma}}\, .
\een
For the fragmentation photons, there are both Sivers and Collins contributions for the spin asymmetry, we thus further define the Sivers asymmetry for fragmentation photons $A_{N, \rm Sivers}^{\rm frag}$, and the Collins asymmetry for fragmentation photons $A_{N, \rm Collins}^{\rm frag}$,
\ben
A_{N, \rm Sivers}^{\rm frag} = \frac{E_\gamma\frac{d\Delta\sigma^{\rm frag}_{\rm Sivers}}{d^3P_\gamma}}{E_\gamma\frac{d\sigma^{\rm frag}}{d^3P_\gamma}},
\qquad
A_{N, \rm Collins}^{\rm frag} = \frac{E_\gamma\frac{d\Delta\sigma^{\rm frag}_{\rm Collins}}{d^3P_\gamma}}{E_\gamma\frac{d\sigma^{\rm frag}}{d^3P_\gamma}}\, .
\een
Note that  the spin asymmetry for fragmentation photons is the sum of Sivers and Collins asymmetry as
$A_N^{\rm frag} = A_{N, \rm Sivers}^{\rm frag}+A_{N, \rm Collins}^{\rm frag}$. However, the overall spin asymmetry for prompt photon $A_N \neq A_N^{\rm dir} + A_N^{\rm frag}$.

The quark-to-photon fragmentation function has been extracted from the phenomenological study, see e.g.,  Ref.~\cite{Gluck:1992zx}. This parametrization has been used to describe the unpolarized prompt photon production at RHIC energy \cite{Adare:2012yt}. Thus,  to compute the fragmentation photon cross section in spin-averaged proton-proton collisions, we will use this phenomenological parametrization instead of the model result in Eq.~(\ref{Dq}). On the other hand, since there is no experimental information at all for the quark-to-photon twist-three fragmentation function $\hat{H}_q(z)$, we  rely on our model calculation in order to estimate the Collins contribution to the asymmetry of fragmentation photons. 
In this case, we will assume that our model calculations give a reasonable estimate on the relative size for $\hat{H}_q(z, \mu^2)$ and $D_{q\to \gamma}(z, \mu^2)$. Thus we will use the following approximation, 
\ben
\left.\frac{\hat{H}_q(z, \mu^2)}{D_{q\to \gamma}(z, \mu^2)}\right|_{\rm phenomenology}
=\left.\frac{\hat{H}_q(z, \mu^2)}{D_{q\to \gamma}(z, \mu^2)}\right|_{\rm model},
\een
where $\hat{H}_q(z, \mu^2)$ and $D_{q\to \gamma}(z, \mu^2)$ on the right-hand side are given by the expressions in Eqs.~(\ref{Dq}) and (\ref{hqcoll}) in our model calculations, $D_{q\to \gamma}(z, \mu^2)$ on the left-hand side is the phenomenological parametrization from Ref.~\cite{Gluck:1992zx}, and $\hat{H}_q(z, \mu^2)$ in the numerator on the left-hand side will be the quark-to-photon twist-three fragmentation function to be used in our calculation for the asymmetry $A_{N, \rm Collins}^{\rm frag}$ of fragmentation photons. For quark transversity distribution  $h_a(x)$, we take the parametrization from Ref.~\cite{Anselmino:2007fs}.

On the other hand, to calculate $A_N^{\rm dir}$ and $A_{N, \rm Sivers}^{\rm frag}$, we need the twist-three quark-gluon correlation functions $T_{q, F}(x, x)$. This function has been extracted directly from the inclusive hadron production in proton-proton collisions \cite{Kouvaris:2006zy}, which will be labeled as ``KQVY'' parametrization in our plots. $T_{q, F}(x, x)$ can also be computed indirectly from Eq.~(\ref{TF}) with the quark Sivers function extracted from SIDIS process \cite{Anselmino:2005ea,Anselmino:2008sga}. Such indirectly obtained parametrization for $T_{q, F}(x, x)$ from \cite{Anselmino:2005ea} will be called ``old'' parametrization, while that from  \cite{Anselmino:2008sga} will be labeled as ``new'' parametrization in our plots. It has been found in \cite{Kang:2011hk} that the {\it directly} and {\it indirectly} obtained $T_{q, F}(x, x)$ 
have conflicting signs, for both $u$ and $d$ quark flavors. The future prompt photon production hopefully could help us pin down the sign and magnitude of $T_{q, F}(x, x)$.

\bef
\psfig{file=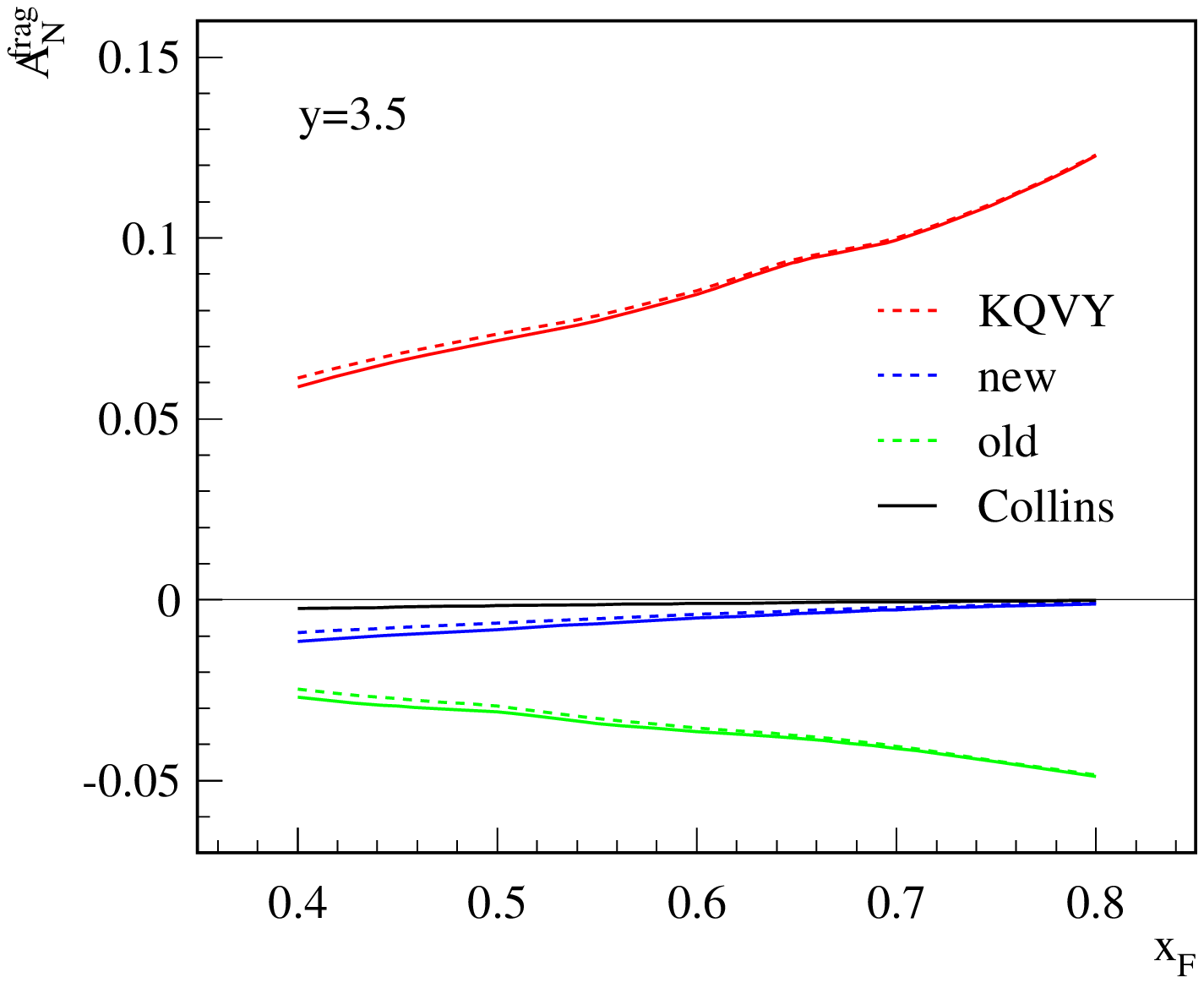, width=3.0in}
\hskip 0.3in
\psfig{file=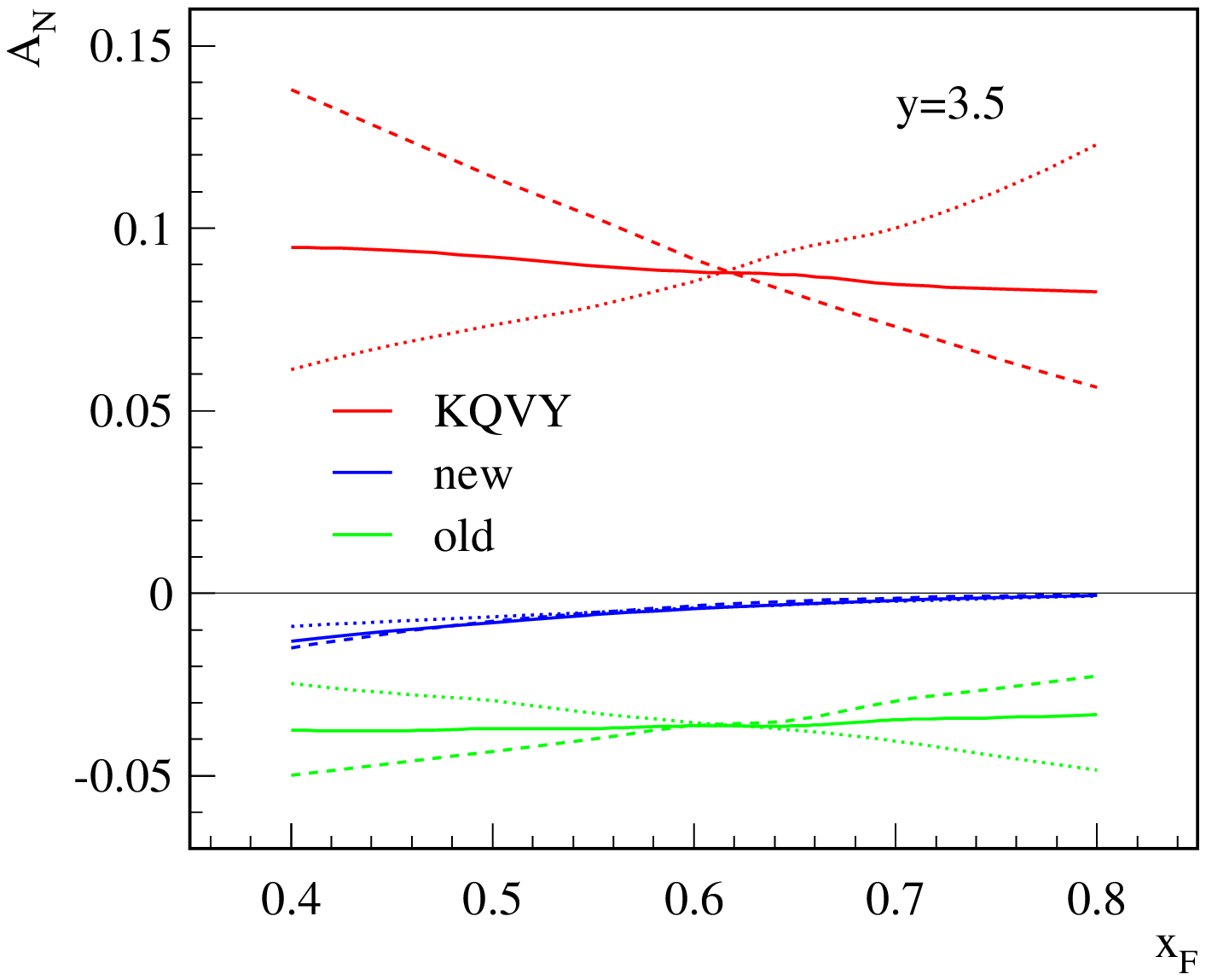, width=3.0in}
\caption{Single transverse spin asymmetry for prompt photon production, $p^\uparrow+p\to \gamma+X$,
is plotted as a function of Feynman $x_F$ at rapidity $y=3.5$ and center-of-mass energy $\sqrt{s}=200$ GeV.
Left panel: the asymmetry for the fragmentation photons. The black solid curve is the Collins asymmetry $A_{N, \rm Collins}^{\rm frag}$. Dashed curves are the Sivers asymmetry $A_{N, \rm Sivers}^{\rm frag}$, with the red curve for ``KQVY'' parametrization, the blue curve for ``new'' parametrization, and the green curve for ``old'' parametrization for $T_{q, F}(x, x)$. For each set, the solid curve is the asymmetry for fragmentation photons $A_{N}^{\rm frag}$, which is the sum of $A_{N, \rm Collins}^{\rm frag}$ and $A_{N, \rm Sivers}^{\rm frag}$. 
Right panel: the asymmetry for the prompt photons. For each set, the dashed curve is the direct asymmetry $A_N^{\rm dir}$, the dotted curve is the fragmentation asymmetry $A_{N}^{\rm frag}$, and the solid curve is the overall spin asymmetry $A_N$.}
\label{AN}
\eef

In Fig.~\ref{AN}(left), we plot the spin asymmetry for fragmentation photons as a function of Feynman $x_F$ at forward rapidity $y=3.5$ and RHIC energy $\sqrt{s}=200$ GeV. The black solid curve is the Collins asymmetry $A_{N, \rm Collins}^{\rm frag}$. We find that the Collins asymmetry is very small in the whole $x_F$ region, less than $1\%$.  Dashed curves are the Sivers asymmetry $A_{N, \rm Sivers}^{\rm frag}$, with the red curve for ``KQVY'' parametrization, the blue curve for ``new'' parametrization, and the green curve for ``old'' parametrization for $T_{q, F}(x, x)$. For each set, the solid curve is the asymmetry for fragmentation photons $A_{N}^{\rm frag}$, which is the sum of $A_{N, \rm Collins}^{\rm frag}$ and $A_{N, \rm Sivers}^{\rm frag}$. In Fig.~\ref{AN}(right), we plot the spin asymmetry for the prompt photons. For each set, the dashed curve is the direct asymmetry $A_N^{\rm dir}$, the dotted curve is the fragmentation asymmetry $A_{N}^{\rm frag}$, and the solid curve is the overall spin asymmetry $A_N$.
We find that the spin asymmetry for fragmentation photons $A_{N}^{\rm frag}$ has the same sign as the direct asymmetry $A_N^{\rm dir}$, thus the overall spin asymmetry $A_{N}$ has the same sign as $A_N^{\rm dir}$ and $A_N^{\rm frag}$.

Some  comments are in order on the reliability our model estimate 
of the photon Collins contribution in prompt photon production. 
We re-emphasize, in this partonic model picture  
the quark-to-photon Collins  function is set by the electro-magnetic 
and strong  couplings, as well as the chiral-symmetry breaking quark 
mass~\cite{Bacchetta:2002tk,Amrath:2005gv,Bacchetta:2007wc}. 
Further,  fixing the quark mass by making a best estimate to 
phenomenological extraction of the   
unpolarized photon fragmentation function, 
we then find that the photon Collins contribution is relatively small.  
While this estimate of the Collins effect is derived from a specific model 
calculation we expect this behavior from any partonic description of 
the photon Collins function.   Thus, within this partonic framework the 
Collins asymmetry for fragmentation photons is very small, and  the asymmetry of prompt photon production can possibly be a very good probe for the twist-three quark-gluon correlation functions $T_{q, F}(x, x)$.
We urge the experiments to measure the asymmetry of prompt photon production at RHIC. It will provide important information on the twist-three quark-gluon correlation functions, a quantity much needed to verify our current theoretical formalism for describing single transverse spin asymmetry in proton-proton scatterings. The measurement can go a long
way to resolving the so called ``sign mismatch'' \cite{Kang:2011hk,Kang:2012xf,Gamberg:2010tj}.

\section{Summary}
We have studied the single transverse spin asymmetry of prompt photon production in high energy proton-proton scattering including the contributions from both the direct and fragmentation photons. While the
asymmetry for direct photon production receives only the Sivers type of contribution, the asymmetry for fragmentation photons receives both the Sivers and Collins types of contributions. In order to estimate 
the Collins asymmetry for fragmentation photons, we perform a model calculation for the chiral-odd quark-to-photon 
Collins function.   Our estimate of the Collins asymmetry is derived from 
a  partonic model calculation extended from that for  quark-to-pion 
fragmentation~\cite{Metz:2002iz,Gamberg:2003eg,Amrath:2005gv,Bacchetta:2007wc}.
In order to obtain a non-trivial Collins effect in  this framework 
we estimate the chiral-odd property of the Collins effect by choosing a non-zero
quark mass of $m_q=300 \ {\rm MeV}$. This framework  has been shown to give  reasonable estimate of unpolarized  quark-to-photon fragmentation function. 
Further based on the
fundamental quark-photon and quark-gluon interactions 
we expect  it  characterizes  
the dynamics of the photon Collins function and in 
turn yields a reasonable estimate of the photon Collins contribution to the
prompt photon production. We find that the Collins asymmetry for fragmentation 
photons is very small in the whole kinematic region, thus the single transverse spin asymmetry of prompt 
photon production is mainly coming from the Sivers asymmetry in direct and fragmentation photons.
We hope the experiments in the future could constrain the different contributions to the prompt photon production, e.g., through an isolation cut.
We further make predictions for the prompt photon spin asymmetry at RHIC energy, and find that  the asymmetry is sizable. The asymmetry of prompt photon production should then provide a good measurement for the important twist-three quark-gluon correlation function, which is urgently needed in order to resolve the ``sign mismatch'' puzzle. We urge the experiments to measure the asymmetry of prompt photon production at RHIC in the near future. 

\begin{acknowledgments}
We thank W.~Vogelsang for providing us the NLO code used to calculate the ratio in Fig.~\ref{figratio}, and thank L.~Eun, X.~Jiang, M.~Liu, R.~Seto, A.~Vossen, and I.~Younus for useful discussions on the experimental measurements. This work was supported in part by the U.S. Department of Energy under Contract 
Nos. DE-FG02-07ER41460 (L.G.)  and DE-AC02-05CH11231 (Z.K.).
\end{acknowledgments}


\end{document}